# Reply to "Comment on `Theory for tailoring sonic devices: Diffraction dominates over refraction' "

N.Garcia[1], M.Nieto-Vesperinas[2], E.V.Ponizovskaya[1] and M.Torres[3]

[1] Laboratorio de Fisica de Sistemas Pequeños y Nanotecnologia, Consejo Superior de Investigaciones Cientificas, Serrano 144, 28006 Madrid, Spain

[2] Instituto de Ciencia de Materiales de Madrid, Consejo Superior de Investigaciones Cientificas, Campus de Cantoblanco Madrid 28049 Madrid, Spain

[3] Instituto de Fisica Aplicada, Consejo Superior de Investigaciones Cientificas, Serrano 144, 28006 Madrid, Spain

In their comment, A. Hakansson, J. Sanchez-Dehesa, F. Cervera, F. Meseguer, L. Sanchis, and J. Llinares say that our conclusion stating that diffraction prevails over refraction in acoustic lenses whose aperture is of several wavelengths, such as those addressed in our calculations [N. Garcia, M. Nieto-Vesperinas, E. V. Ponizovskaya, and M.Torres, Phys. Rev. E 67, 046606 (2003)] and in their experiments [F.Cervera, L. Sanchis, J. V. Sanchez-Perez, R. Martinez-Sala, C. Rubio, F.Meseguer, C. Lopez, D. Caballero, and J. Sanchez-Dehesa, Phys. Rev. Lett. 88, 023902 (2003)], is misleading because the size of their lenses is larger than ours. They state that diffraction effects are negligible at the scale of their experiments. In this reply we calculate the propagation of a plane wave through both a lens and a slab of aluminum cylinders, identical to those presented by such authors in previous experiments, by using a finite difference time domain (FDTD) method. We then compare our results to the experiments previously reported by the authors of the comment and significant differences are found. Our present calculations show that refraction and diffraction are intrinsically interwoven also at the scale of their experiments.

For the acoustic devices under discussion, refraction is dominant upon diffraction if both their aperture, (namely, their lateral size), and the wavelength versus the lattice constant are large enough, so that edge effects are minimal, and an effective medium can be identified. However, diffraction becomes more and more dominant as the situation progressively deviates from the above conditions. However, when such conditions are not fully satisfied, like when the aperture is only about some wavelengths, and the lattice constant is not much smaller than the wavelength, one should expect an intermediate regime in which diffraction coexists with refraction.

Reference [1] unambiguously proved that acoustic lenses with size of the order of the wavelength can be realized; demonstrating not only focusing, but image formation (see Fig. 1 and 2 of Ref. [1]). The authors of the comment agree that in such lenses "focusing phenomena and image formation are dominated by diffraction rather than refraction due to the small dimension of the acoustic devices studied". However, the size of the lens used by the authors of the comment is about six wavelengths and the difference between one and six wavelength may be crucial and this should be investigated.

To do that, we calculate here, by using an FDTD method described elsewhere [1,3,4], the propagation through acoustical devices identical to those presented by the authors of the comment in their previous experiments [2] As it is well known, the FDTD method allows one to simulate actual experiments. The author of the comment also make our calculations, obtaining a result that does not fully agree with ours, although they point out a similarity. We believe, however, that the calculation in that comment has not enough accuracy, because we have performed our computations by doubling the number of points, and have obtained the same result as before. (In fact, the same method had been previously checked in analogous calculations presented together with M. Kafesaki and M. M. Sigalas some years ago [3, 4]). Furthermore, we believe that the experiments addressed both in the comment and in Ref. [2], are not accurate because they are not scaled with the curvature of the lens, and this leads to something like a theory of universal focus: no matter what the lens size or geometry is, the same focus position is obtained, or so it is concluded after reading the comment.

To show that diffraction is not negligible in previous experiments [2] of the authors of the comment, we present here the result of a new numerical experiment, developed according to the aforementioned FDTD method. Both a sonic lens, identical to that presented in [2], and a sonic crystal slab with parallel faces made of 10 monolayers of aluminum cylinders (49.5 cm thickness) oriented along the $\Gamma X$ direction in the hexagonal structure, (this is exactly the same as in Ref. [2], Fig. 4 (bottom and top respectively)), are illuminated with an incident sound plane wave in air at 1700 Hz. The corresponding intensity patterns are shown in Fig. 1. The pattern corresponding to the lens (Fig. 1 (a)) shows a conspicuous maximum just at the right lens apex in the symmetry axis of the system. This maximum corresponds to the stronger intensity at the right side of the acoustic lens (0 dB, i.e., the same intensity as the incident plane wave) but it is unnoticed in recent experiments [2]. The physical origin of this intensity peak cannot be associated to any refractive phenomenon of an effective medium as that reported in reference [2]. On the other hand, there is a sort of focus in the symmetry axis at 86 cm away from the apex of the lens, but it is not clear because for a true focus one would expect its intensity to be stronger than anywhere else behind the lens, but in this case there are other regions with similar intensity. In fact, other sort of focusing appears 136 cm from the lens apex with exactly the same intensity as the former, as well as other two peaks with the same intensity, both above and below the symmetry axis respectively. In any case, the general geometrical complexity of the intensity pattern shown in Fig. 1 (a), does not fit well the experimental result reported in reference [2]. Our theoretical results suggest that both refraction and diffraction phenomena are intrinsically mixed in this wave propagation régime.

On the other hand, in Fig. 1 (b) we show the intensity pattern behind the slab of aluminum cylinders. This intensity distribution at the right of the sonic slab, constitutes an interference pattern (also unnoticed in Ref. [2]) resulting from two wavefronts propagating along the ΓJ main directions in the hexagonal structure, and emerging at its right side along two directions at $60^o$ with each
other. Taking into account that there is normal incidence of a plane wave, the interference behind the slab cannot be associated to any refraction phenomenon of an effective medium, but it should correspond to the directions that give rise to two diffraction orders from this two dimensional diffraction grating constituted by the cylinder array. These directions are represented in Fig. 1 (b). The intensity pattern is slightly asymmetrical about the horizontal line due to the asymmetry at the edges of the slab, exactly the same as the experiments reported in Ref. [2]. This indicates the great importance of the diffraction effects present at the edges zones. Furthermore, the anomalous refraction due to the anisotropy of the isofrequency k-curves should be also discarded because the frequency of the experiment is well below the gap and it is well known that the phenomena of anomalous or negative refraction can only be produced at frequencies placed just at the exit of the first gap at the boundary of the first Brillouin zone [5].

As a main conclusion of this reply, we infer from our present calculations that both refraction and diffraction phenomena are inextricably entwined in experiments reported in Ref. [2].

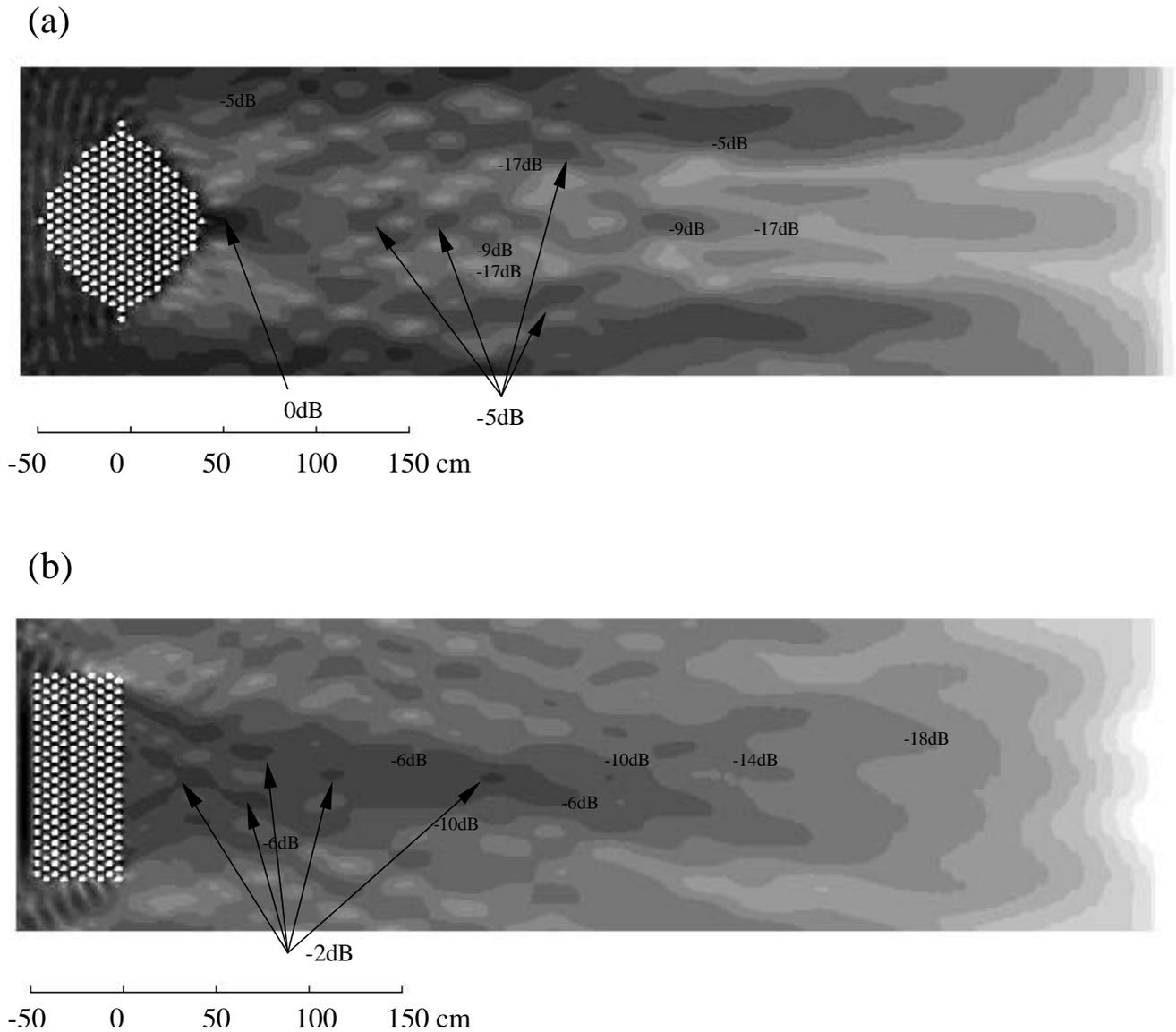

Fig.1 Intensity pattern of an incident sound plane wave in air, at 1700 Hz, illuminating a sonic lens (a) and a crystal slab made of 10 monolayers of aluminum cylinders oriented along the ΓX direction in the hexagonal structure (b). There is no clear focus behind the sonic lens but four maxima with an intensity of –5 dB and a conspicuous maximum with intensity of 0 dB just at the apex of the lens that can not be explained by any refraction phenomenon (a). The interference pattern at the right side of the sonic slab can not be attributed to two refracted waves because the source placed at the left of the slab is a plane wave and this is a case of normal incidence (b).